# Thick lenses systems

Fulvio Andres Callegari

Centro de Engenharia, Modelagem, e Ciências Sociais Aplicadas, Universidade Federal do ABC, Santo André, SP, Brasil. Avenida dos Estados, 5001 - Bairro Santa Terezinha Santo André - CEP: 09210-580

**Corresponding Author: fulvio.callegari@ufabc.edu.br**

**Abstract**
The optical matrix formalism is applied to find parameters such as focal distance, back and front focal points, principal planes, and the equation relating object and image distances for a thick spherical lens immerse in air. Then, the formalism is applied to systems compound of two, three and *N* thick lenses in cascade. It is found that a simple Gaussian equation is enough to relate object and image distances no matter the number of lenses.

**Keywords:** Matrix Optics, Thick Lenses, Back Focal length, Front Focal length, Principal Planes, Focus, Multi-lenses system.

## I. INTRODUCTION

The optical matrix formalism is a powerful tool that can be applied to the study of thick multi-lenses systems. It allows an intuitive approach in order to understand more deeply dispositive such as cameras, that may consist of several thick lenses in cascade. By the order side, when the object-image equation for a single thick lens is deduced using matrix formalism, it is obtained a complicated expression to deal with. However, via a coordinate transformation on the image and object distances, it is possible to obtain a very interesting result which simplifies the expression to a formula that is formally identic to the familiar expression used for thin lenses i.e. the Gaussian equation. In this work, I fully derive such expression introducing optical parameters such as effective focal distance, front and back frontal length, and principal planes. This is not a new result, as it has been pointed out in [1-3]. Nevertheless, in this work I demonstrate this feature in a different way, by using the matrix formalism (not used in the references before mentioned for this specific result), which allows to explore other situations. Then, I extended this analysis to a system formed by two thick lenses separated by a distance *d*. It was found the interesting result that all optical parameters characterizing this system and the object-image equation are mathematically identical to the previously found expressions for a single thick lens. Furthermore, I extended this analysis to optical systems consisting in three and *N* thick lenses, and formally identical results were found once again. Indeed, one of the more interesting results in this work is this generalization and, as a direct consequence, the possibility of expressing the relation between object and image distances for an arbitrary number of thick lenses in cascade just in a single equation. Actually, this general result is predicted by Feynman [4], without demonstration.

## II. MATRIX OPTICS

The optical matrices used for this study, i.e., the refraction and the displacement matrices, can be found in [1]. An important remark is that these matrixes are deduced for the paraxial approximation, then, all our results are valid in that case. Here, I will briefly derive the results. Then, the interpretation and further application of these matrices to develop the expressions for spherical thick lenses are shown. Also I adopted the classical signal convention, which states that object distances to the left (right) of some reference interface will be positive (negative), image distances place to the right (left) of such interface will be positive (negative). As for the curvature radios of surfaces, they will be positive (negative), if they were convex (concave). This convention is currently adopted in practically all specialized literature.

Let´s suppose an optical ray that makes an angle $\alpha_1$ with respect to horizontal direction *z* (from here on, the optical axis of the system), is propagated in an optical medium with a refraction index $n_1$. Then, it hits a spherical medium with refraction index $n_2$ at a height *y* on the surface, measure from the optical axis, suffers refraction and, as a consequence, changes the angle with respect to the optical axes to $\alpha_2$, see Fig. 1.

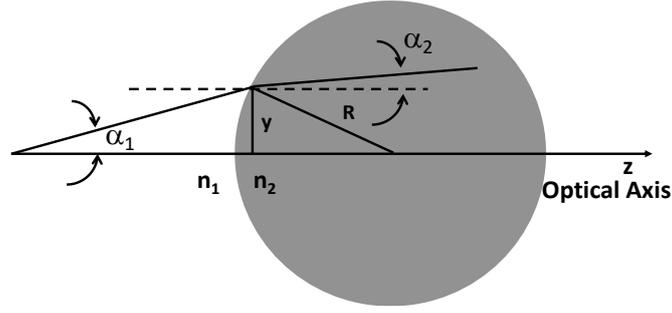

**Figure 1:** Refraction of an optical ray in a spherical surface.

In Fig.1, 'R' denotes the radium of the sphere. The equation that completely describes refraction is:

$$n_2 \alpha_2 = n_1 \alpha_1 + y \frac{n_1 - n_2}{R}. \tag{1}$$

Note that in Eq. (1), if 'R' tends to infinity, then the equation reduces to Snell law (in paraxial form), as expected. We can express Eq. (1) in a matrix form, by defining the vector ($n\alpha$, $y$):

$$\begin{pmatrix} n_2 \alpha_2 \\ y_2 \end{pmatrix} = \begin{pmatrix} 1 & -\dfrac{n_2 - n_1}{R} \\ 0 & 1 \end{pmatrix} \begin{pmatrix} n_1 \alpha_1 \\ y_1 \end{pmatrix} \tag{2}$$

Applying the usual operation of matrix product, the result corresponding to the $n_2\alpha_2$ component reproduces Eq. (1). The other term, i.e., $y_2 = y_1$, simply describes that no change in height was verified at the point of refraction.

Now we describe the change of height, i.e., in '$y$' coordinate of the optical ray, as it propagates in an optical medium. See Fig. 2.

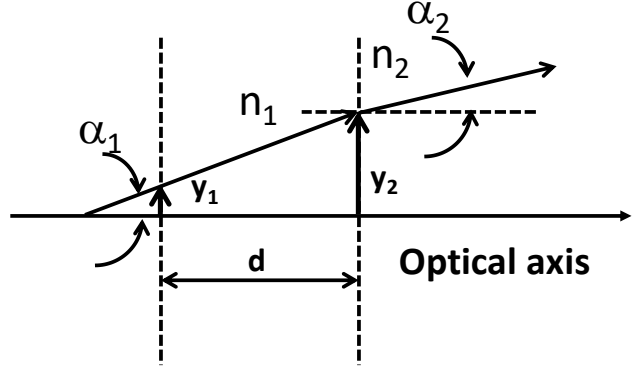

**Figure 2:** Propagation of an optical ray in an optical medium.

Is easy to see that, as the ray propagates a distance *d* measured over the optical axis in the medium with refraction index $n_1$, the changes in $y$ coordinate is given by:

$$y_2 = y_1 + d \tan \alpha_1. \tag{3}$$

As we work with the paraxial approximation, i.e., $\alpha_1 \ll 1$, so, $\tan \alpha_1 \sim \alpha_1$, Eq.(3) becomes:

$$y_2 = y_1 + d \alpha_1. \tag{4}$$

Now, we can write Eq. (4) in matrix form as:

$$\begin{pmatrix} n_2 \alpha_2 \\ y_2 \end{pmatrix} = \begin{pmatrix} 1 & 0 \\ \dfrac{d}{n_1} & 1 \end{pmatrix} \begin{pmatrix} n_1 \alpha_1 \\ y_1 \end{pmatrix}. \qquad (5)$$

Performing the product, the expression that corresponds to the element '$y_2$', reproduces Eq.(4), while the result corresponding to the component $n_2\alpha_2$ reproduces the Snell law, in paraxial form, at the interface between the two medium.

## III. THICK LENSES IN AIR, FOCAL POINTS AND PRINCIPAL PLANES.

The vantage of the matrix formalism is that it can be applied to several optical mediums in cascade by simply multiplying the matrixes corresponding to each element. Now, it will be applied to a spherical thick lens immersed in air, which consist in two refractive spherical surfaces of radios $R_1$ and $R_2$, separated a distance *d* over the optical axis, which is the thickness of the lens, as shown in Fig. 3.

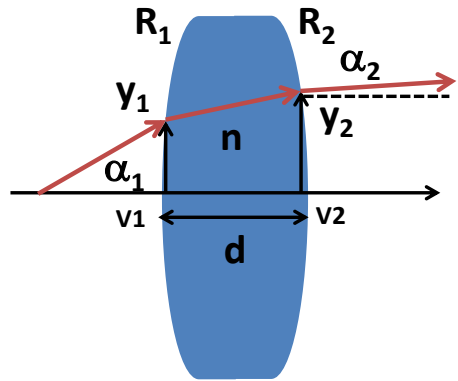

**Figure 3:** Thick lens.

In Fig. 3, $V_1$ and $V_2$ are vertices points, localized at the intersections of the spherical surfaces of radio $R_1$ and $R_2$, with the optical axis, respectively. Also, it can be seen that an optical ray making an $\alpha_1$ angle with respect to the optical axis is refracted by the first surface, at a height $y_1$, propagates inside the lens (a distance *d* measured over the optical axis) and then is refracted by the second surface, leaving the lens with an angle of $\alpha_2$ at a height $y_2$ with respect to the optical axis. In order to mathematically describe the journey of the optical ray through the lens it can be written:

$$\begin{pmatrix} \alpha_2 \\ y_2 \end{pmatrix} = \begin{pmatrix} 1 & -\dfrac{1-n}{R_2} \\ 0 & 1 \end{pmatrix} \begin{pmatrix} 1 & 0 \\ \dfrac{d}{n} & 1 \end{pmatrix} \begin{pmatrix} 1 & -\dfrac{n-1}{R_1} \\ 0 & 1 \end{pmatrix} \begin{pmatrix} \alpha_1 \\ y_1 \end{pmatrix}. \qquad (6)$$

As the lens is immersed in air, the indexes of refraction that multiply the angles in the vector components are both equal to one. Also, the elements corresponding to file 1 column 2 in both of the refraction matrixes in Eq. (6) can be easily understood by looking to the general form of this element, in Eq. (2), it's numerator could be described in words as: "(minus) index of refraction to the right of the interface minus the index of refraction to the left of the interface (over R)". Then, the first interface (with radius '$R_1$') is surrounded by air (index $n_{ar}=1$) to the left and glass (of index *n*) to the right. The same applies to the second surface (with radius '$R_2$').

A convenient way to look at Eq. (6) is from right to left: first, we have the input ray, hitting the lens with angle $\alpha_1$ at a height $y_1$, it is refracted by the first surface, with radio $R_1$, then the ray is displaced a horizontal distance *d*, inside the lens, which has a refraction index *n*, and finally it is refracted by the second refractive surface, with radius $R_2$, and this gave the resultant optical ray leaving the lens, characterized by $\alpha_2$ and $y_2$. Recalling that matrix product is not commutative, the correct order of the matrixes is fundamental in order to achieve the right result.

Performing the usual operation of matrix product, Eq. (6) results:

$$\begin{pmatrix} \alpha_2 \\ y_2 \end{pmatrix} = \begin{pmatrix} 1 + \dfrac{n-1}{n}\dfrac{d}{R_2}, & -\left[(n-1)\left(\dfrac{1}{R_1} - \dfrac{1}{R_2}\right) + \dfrac{(n-1)^2 d}{nR_1 R_2}\right] \\ \dfrac{d}{n}, & 1 - (n-1)\dfrac{d}{nR_1} \end{pmatrix} \begin{pmatrix} \alpha_1 \\ y_1 \end{pmatrix}. \qquad (7)$$

Note that, if $d = 0$, i.e., when the thick lens becomes a thin lens, the element $a_{12}$ in the matrix characterizing the lens, becomes equal to minus the inverse of the focal distance for a thin lens. From here on, I will write this term as -$1/f$ (parameter $f$ for thick lens will be analyzed later on in this section).

Now I will deduce the back focal length, $z_b$, that is the distance, measured from $V_2$, to which a thick lens focalizes a bunch of optical ray incident parallel to the optical axis, see Fig. 4.

Equivalently, it may be though as a plane wave front (being the wave fronts perpendicular to the optical rays) that is incident in the lens, and, after refraction, a convergent spherical wave emerges, whose center is at back focal point. Some of these fronts are shown in Fig. 4.

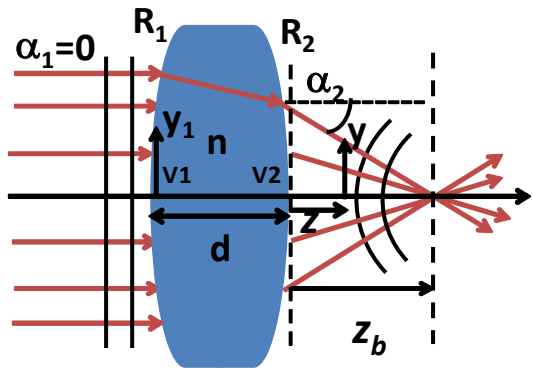

**Figure 4:** Finding back focal length for a thick lens.

Now, I will write the expression to find the height $y$, over the optical axis, of an optical ray that has travelled a horizontal distance $z$ after leaving the lens. In order to mathematically describe the whole process, it must be used, in addition to the thick lens matrix, the displacement matrix in air in the following way:

$$\begin{pmatrix} \alpha_2 \\ y \end{pmatrix} = \begin{pmatrix} 1 & 0 \\ z & 1 \end{pmatrix} \begin{pmatrix} 1 + \dfrac{n-1}{n}\dfrac{d}{R_2} & -\dfrac{1}{f} \\ \dfrac{d}{n} & 1 - (n-1)\dfrac{d}{nR_1} \end{pmatrix} \begin{pmatrix} \alpha_1 \\ y_1 \end{pmatrix}. \qquad (8)$$

Again, it´s convenient to read Eq. (8) from right to left, i.e., an input ray ($\alpha_1$ $y_1$) is refracted by the lens, and then propagates an horizontal distance $z$. Performing the product of the matrixes, Eq. (8) becomes:

$$\begin{pmatrix} \alpha_2 \\ y \end{pmatrix} = \begin{pmatrix} 1 + \dfrac{n-1}{n}\dfrac{d}{R_2}, & -\dfrac{1}{f} \\ \left(1 + \dfrac{n-1}{n}\dfrac{d}{R_2}\right)z + \dfrac{d}{n}, & 1 - (n-1)\dfrac{d}{nR_1} - \dfrac{z}{f} \end{pmatrix} \begin{pmatrix} \alpha_1 \\ y_1 \end{pmatrix}. \qquad (9)$$

Now, the expression for $y$:

$$y = \left[\left(1 + \frac{n-1}{n}\frac{d}{R_2}\right)z + \frac{d}{n}\right]\alpha_1 + \left[1 - (n-1)\frac{d}{nR_1} - \frac{z}{f}\right]y_1. \tag{10}$$

Noting that in our particular case $\alpha_1=0$, we seek for that value of $z$ that makes $y$ null for all values of $y_1$ (see Fig. 4). This is, by definition, the back focal point, $z_b$:

$$z_b = f\left[1 - (n-1)\frac{d}{nR_1}\right]. \tag{11}$$

Note that, if $d=0$, $z_b=f$, i.e., the same result for thin lenses, as expected (and in this particular case, of course, the focal distance became the familiar expression for thin lenses).

A consideration of signal must be mentioned here. Back focal length is, obviously, an image point, and the corresponding signal convention should be applied, i.e. if the result is positive, it should be localized to the right of $V_2$, as schematically shown at Fig. (4), where a positive value of $z_b$ is supposed. If negative, it should be localized to the left of $V_2$.

Now it can be deduced the front focal length, $z_f$, that is the distance, measure from $V_1$ on the optical axis, in which an object point (or a source of spherical divergent wave fronts), must be located in order that a thick lens produces a bunch of parallel rays (corresponding to plane wave fronts) after refraction. The situation is shown in Fig. 5.

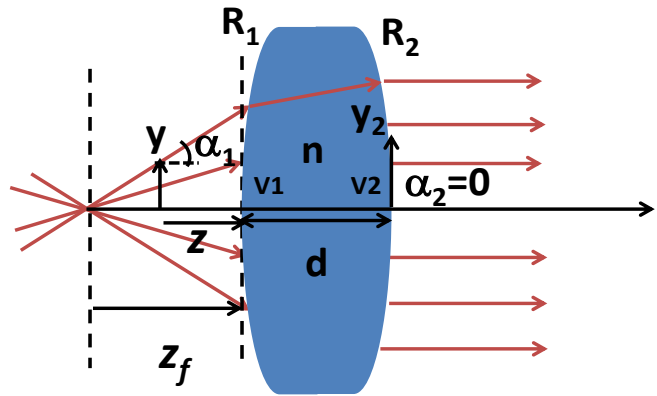

**Figure 5:** Finding front focal length for a thick lens.

Now we write the expression to find $(\alpha_2, y_2)$:

$$\begin{pmatrix}\alpha_2 \\ y_2\end{pmatrix} = \begin{pmatrix} 1 + \frac{n-1}{n}\frac{d}{R_2} & -\frac{1}{f} \\ \frac{d}{n} & 1-(n-1)\frac{d}{nR_1} \end{pmatrix}\begin{pmatrix}1 & 0 \\ z & 1\end{pmatrix}\begin{pmatrix}\alpha_1 \\ y\end{pmatrix}. \tag{12}$$

Note the order in the matrix's product in Eq. (12), indicating that a ray with height $y$ and angle $\alpha_1$ is propagated a certain distance $z$, before it hits the lens. Performing the product indicated in Eq. (12):

$$\begin{pmatrix}\alpha_2 \\ y_2\end{pmatrix} = \begin{pmatrix} 1 + \frac{n-1}{n}\frac{d}{R_2} - \frac{z}{f} & -\frac{1}{f} \\ \frac{d}{n} + \left(1-(n-1)\frac{d}{nR_1}\right)z & 1-(n-1)\frac{d}{nR_1} \end{pmatrix}\begin{pmatrix}\alpha_1 \\ y\end{pmatrix}. \tag{13}$$

Now, we analyze the term corresponding to $\alpha_2$ from Eq.(13):

$$\alpha_2 = \left(1 + \frac{n-1}{n}\frac{d}{R_2} - \frac{z}{f}\right)\alpha_1 - \frac{y}{f}. \tag{14}$$

We must impose the conditions $y = 0$, and $\alpha_2 = 0$, which are compatible with an object point over the optical axis producing, via the thick lens, an "image at infinity" (or a bunch of parallel rays), as indicated in Fig. 5. As a consequence, the term between parentheses in Eq. (14) should be equal to zero, for any value of $\alpha_1$. The particular value of $z$ that makes this condition be fulfilled is the $z_f$:

$$z_f = f\left[1 + (n-1)\frac{d}{nR_2}\right]. \tag{15}$$

If we set $d = 0$, i.e., a thin lens, Eq. (15) reduces, as expected, to the expression for thin lenses.

Front focal point is an object point and, if positive, it should be located to the left of $V_1$ (as indicated in Fig. 5, in where that case was supposed). If it has negative signal, of course, it should be localized to the right of $V_1$.

We can now investigate the significance of $f$ for thick lenses.

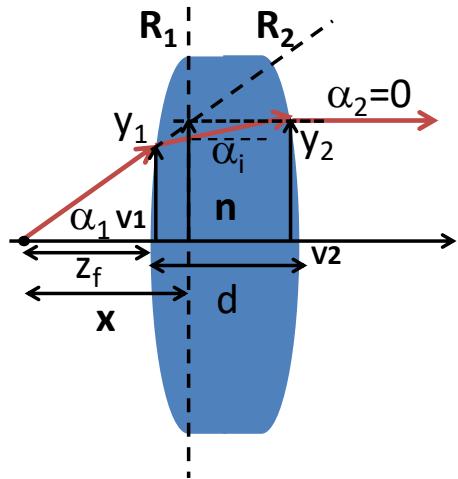

**Figure 6:** Finding $f$ meaning for a thick lens.

In Fig. 6, we see the front focal point for a given thick lens, and a generic optical ray, with $\alpha_1$ angle, entering the lens at $y_1$ and leaving the lens at $y_2$, with $\alpha_2 = 0$ (definition of $z_f$). We now project back the output ray until intersect the projection of input ray. Doing this for every possible $\alpha_1$ angle of those optical rays departing of $z_f$ being refracted by the thick lens, we obtain a plane, perpendicular to the optical axis, at a distance x from $z_f$. This plane is called principal plane. In order to find the distance $x$, we observe that we can express the angle $\alpha_1$, in the paraxial approximation as:

$$\alpha_1 \cong \frac{y_2}{x} \cong \frac{y_1}{z_f}. \tag{16}$$

Then we express $x$ as:

$$x = \frac{y_2}{y_1} z_f. \tag{17}$$

To find $y_2$ we write, from Eq. (7):

$$y_2 = \frac{d\alpha_1}{n} + y_1\left[1 - \frac{(n-1)d}{nR_1}\right]. \tag{18}$$

Again, we must write the expression for $\alpha_1$ in the paraxial approximation:

$$y_2 = \frac{dy_1}{nz_f} + y_1\left[1 - \frac{(n-1)d}{nR_1}\right]. \tag{19}$$

And obtain:

$$\frac{y_2}{y_1} = \frac{d}{nz_f} + 1 - \frac{(n-1)d}{nR_1}. \tag{20}$$

Now, using Eq. (20) we can rewrite Eq. (17) as:

$$x = \frac{y_2 z_f}{y_1} = \frac{d}{n} + z_f\left[1 - \frac{(n-1)d}{nR_1}\right]. \tag{21}$$

Writing explicitly the expression for $z_f$, Eq. (21) becomes:

$$x = \frac{d}{n} + f\left[1 + (n-1)\frac{d}{nR_2}\right]\left[1 - \frac{(n-1)d}{nR_1}\right]. \tag{22}$$

From Eq. (22), and with a little algebra, is easy to found that:

$$x = \frac{d}{n} + f\left[1 - \frac{d}{nf}\right] = f. \tag{23}$$

In this way, we found that the distance from $z_f$ to the principal plane is given by $f$. We can also generate another plane, using the concept of back focal point in a totally symmetric way, and it can be found that the distance between this second plane and $z_b$ is also $f$, see Fig. 7.

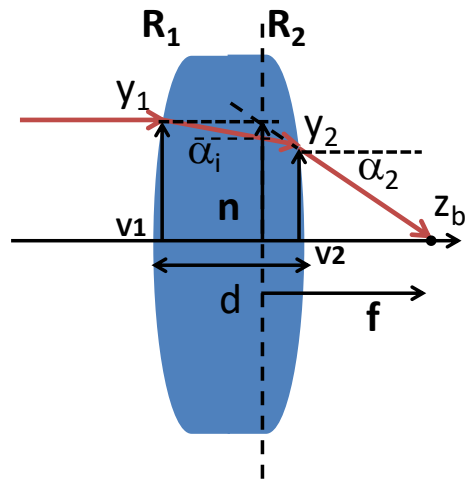

**Figure 7:** Focus and second principal plane for a thick lens.

These two planes are called first principal plane (1PP), and second principal plane (2PP), and they can be seen in Fig. (8).

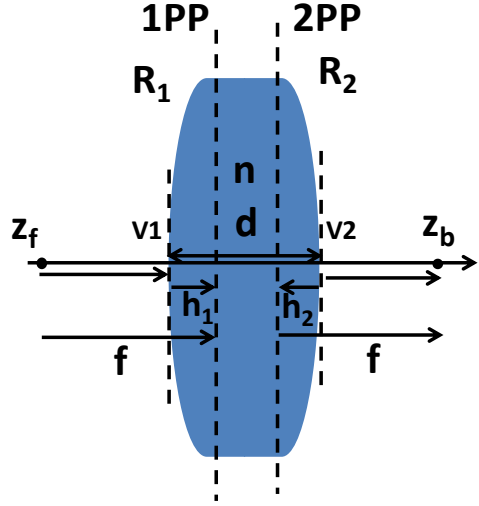

**Figure 8:** Principal planes for thick lenses.

It should be noted that not always the principal planes will be inside the lens, as shown in Fig. 8. Depending on the specific lens data, they could be outside the lens.

For reasons that will be seen in the next section, it is convenient to localize the 1PP and 2PP with respect to the $V_1$ and $V_2$ vertices points, respectively; we call these distances $h_1$ and $h_2$ (see Fig. 8):

$$h_1 = f - z_f = -(n-1)\frac{df}{nR_2}. \qquad (24)$$

$$h_2 = z_b - f = -(n-1)\frac{df}{nR_1}. \qquad (25)$$

From the Eq. (24), it can be seen that distance $h_1$ obeys the object signal convention: if positive, it is located to the right of $V_1$, if negative, to the left of $V_1$. For $h_2$, (see Eq. (25)) the signal convention for images respect to $V_2$ holds, i.e., if positive, $h_2$ must be located to the right of $V_2$ and if negative, to the left of $V_2$.

**IV. THICK LENS IN AIR, IMAGE CONDITION.**

Now, it will be deduced the image condition for a thick lens in air. We consider an object, with height $h_o$ to a distance $s_o$ from $V_1$. An image will be formed by the thick lens at distance $s_i$, measured with respect to $V_2$, and height $h_i$. The situation is visualized in Fig. 9.

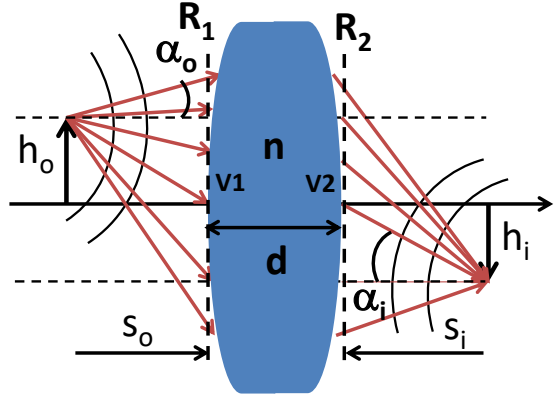

**Figure 9:** Image condition for a thick lens.

We see then that optical rays from object points propagate a distance $s_o$ over the optical axis, then, they will be refracted by the thick lens and propagate a distance $s_i$, to the corresponding image point. In mathematical terms, we can write:

$$\begin{pmatrix} \alpha_i \\ h_i \end{pmatrix} = \begin{pmatrix} 1 & 0 \\ s_i & 1 \end{pmatrix} \begin{pmatrix} a_{11} & a_{12} \\ a_{21} & a_{22} \end{pmatrix} \begin{pmatrix} 1 & 0 \\ s_o & 1 \end{pmatrix} \begin{pmatrix} \alpha_o \\ h_o \end{pmatrix}. \tag{26}$$

Where the matrix *A* represent the thick lens, being its elements:

$$a_{11} = 1 + \frac{n-1}{n}\frac{d}{R_2}. \tag{27}$$

$$a_{12} = -\frac{1}{f} = -\left[(n-1)\left(\frac{1}{R_1} - \frac{1}{R_2}\right) + \frac{(n-1)^2 d}{nR_1 R_2}\right]. \tag{28}$$

$$a_{21} = \frac{d}{n}. \tag{29}$$

$$a_{22} = 1 - (n-1)\frac{d}{nR_1}. \tag{30}$$

Performing the operation indicated in Eq. (26), we arrive to:

$$\begin{pmatrix} \alpha_i \\ h_i \end{pmatrix} = \begin{pmatrix} a_{11} + a_{12} s_o & a_{12} \\ s_i(a_{11} + a_{12} s_o) + a_{21} + a_{22} s_o & s_i a_{12} + a_{22} \end{pmatrix} \begin{pmatrix} \alpha_o \\ h_o \end{pmatrix}. \tag{31}$$

In order to mathematically found the image condition, we can consider an object point, for example, the one at the top of the object. This point is a font of spherical divergent waves (some of these wave fronts are shown in Fig. 9, perpendicular to the optical rays). The lens, in order to form an image, should convert this divergent wave front into a convergent one, whose convergence point is the image point corresponding to the object point at the top of the object. Then, we write, from Eq. (31), the expression corresponding to $h_i$:

$$h_i = [s_i(a_{11} + a_{12}s_o) + a_{21} + a_{22}s_o]\alpha_o + (s_i a_{12} + a_{22})h_o. \tag{32}$$

Is easy to see that the height of a specific image point (for example the one corresponding to the top), should not depend on the angle α₀. In this way, the image condition is:

$$\frac{\partial h_i}{\partial \alpha_o} = 0. \tag{33}$$

Applying Eq. (33) in Eq. (32):

$$s_i(a_{11} + a_{12}s_o) + a_{21} + a_{22}s_o = 0. \tag{34}$$

We explore Eq. (34) when *d*=0, i.e., a thin lens. In that case, the matrix elements become: $a_{11}$ = 1, $a_{12}$ = -1/*f*, where now *f* is reduced to the focal distance for a thin lens, $a_{21}$=0 e $a_{22}$=1. With these values for the matrix elements, is trivial to see that Eq. (34) becomes the well-known Gaussian formula for thin lenses.

We rewrite Eq. (34) as:

$$s_i a_{11} + a_{12} s_o s_i + a_{21} + a_{22} s_o = 0. \tag{35}$$

Now, we seek for new coordinates, *s'ᵢ* and *s'ₒ*, that allows to write Eq. (35) in a simpler way. We can write:

$$s_o = s_o' + o. \tag{36}$$

$$s_i = s_i' + i. \tag{37}$$

Introducing Eqs. (36) and (37) in Eq. (35), and rearranging in a convenient way:

$$s_i'(a_{11} + a_{12}o) + a_{12} s_o' s_i' + s_o'(a_{12}i + a_{22}) + a_{12}oi + a_{21} + a_{22}o + ia_{11} = 0. \tag{38}$$

From Eq. (38) is easy to see that, in order to obtain an expression that looks like the Gaussian formula for thin lens, the following conditions should be imposed:

$$a_{11} + a_{12}o = 1. \tag{39}$$

$$a_{12}i + a_{22} = 1. \tag{40}$$

$$a_{12}oi + a_{21} + a_{22}o + ia_{11} = 0 \tag{41}$$

Solutions to Eqs. (39) and (40) give:

$$o = \frac{1 - a_{11}}{a_{12}}. \tag{42}$$

$$i = \frac{1 - a_{22}}{a_{12}}. \tag{43}$$

Equations (42) and (43) make Eq. (41) be fulfilled (this demonstration requires a certain amount of algebra, and it is shown in the appendix). As a consequence, we were able to find new distances object and image that allows us to write Eq. (35) in a familiar way. Replacing the corresponding elements of the lens matrix in Eq. (42) and Eq. (43), we obtain:

$$o = f(n-1)\frac{d}{nR_2}. \tag{44}$$

$$i = -f(n-1)\frac{d}{nR_1}. \tag{45}$$

These expressions are closely related to Eq. (24) and Eq. (25), i.e., the distances of the principal planes to their respective vertices points. As the distances $s_o$ and $s_i$ are measure with respect to the vertices, this means that, in order to obtain a simpler lens equation, we must measure the distances object and image with respect to the 1PP and 2PP respectively, as shown in Fig. 10:

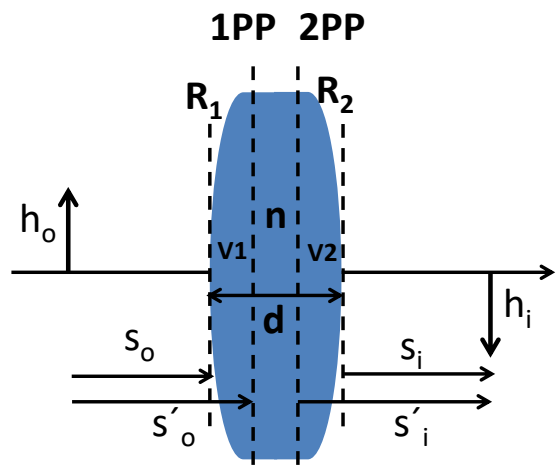

**Figure 10:** Principal planes and image condition for a thick lens.

As a consequence, we can replace the thick lens by the principal planes, and write the Gauss equation:

$$\frac{1}{s'_o} + \frac{1}{s'_i} = \frac{1}{f}. \tag{46}$$

$s'_o$ follows the signal convention for objects; if positive (negative), it is located at the left (right) of 1PP, while $s'_i$ follows the signal convention for images; if positive (negative), it is located at the right (left) of 2PP.

Of course, if we make $d=0$, i.e., a thin lens, the two principal planes coalesce into one, and in that case, $s_o = s'_o$ and $s_i = s'_i$.

**V. TWO THICK LENSES SYSTEMS.**

The formalism developed in the last sections can be generalized to a two lenses system.

Let's suppose a system of two thick lenses, characterized by matrixes *A* and *B*, and separated a distance *d*. The lens *A* has refraction index $n_a$, thickness $d_a$, radius $R_{1a}$, and $R_{2a}$. For the lens *B*, index $n_b$, thickness $d_b$, radius $R_{1b}$, and $R_{2b}$. Matrix elements for each lens are the same as in Eqs. (27) to (30), with the specific parameters for each lens:

$$a_{11} = 1 + \frac{n_a - 1}{n_a} \frac{d_a}{R_{2a}} \quad . \tag{47}$$

$$a_{12} = -\frac{1}{f_a} = -\left[ (n_a - 1)\left(\frac{1}{R_{1a}} - \frac{1}{R_{2a}}\right) + \frac{(n_a - 1)^2 d_a}{n_a R_{1a} R_{2a}} \right]. \tag{48}$$

$$a_{21} = \frac{d_a}{n_a} \quad . \tag{49}$$

$$a_{22} = 1 - \frac{(n_a - 1)}{n_a} \frac{d_a}{R_{1a}} \quad . \tag{50}$$

$$b_{11} = 1 + \frac{(n_b - 1)}{n_b} \frac{d_b}{R_{2b}} \quad . \tag{51}$$

$$b_{12} = -\frac{1}{f_b} = -\left[ (n_b - 1)\left(\frac{1}{R_{1b}} - \frac{1}{R_{2b}}\right) + \frac{(n_b - 1)^2 d_b}{n_b R_{1b} R_{2b}} \right]. \tag{52}$$

$$b_{21} = \frac{d_b}{n_b}. \tag{53}$$

$$b_{22} = 1 - \frac{(n_b - 1)}{n_b} \frac{d_b}{R_{1b}} \quad . \tag{54}$$

The system is shown in Fig. 11. Note that we have now four vertices points, V$_{1a}$, V$_{2a}$, V$_{1b}$, and V$_{2b}$.

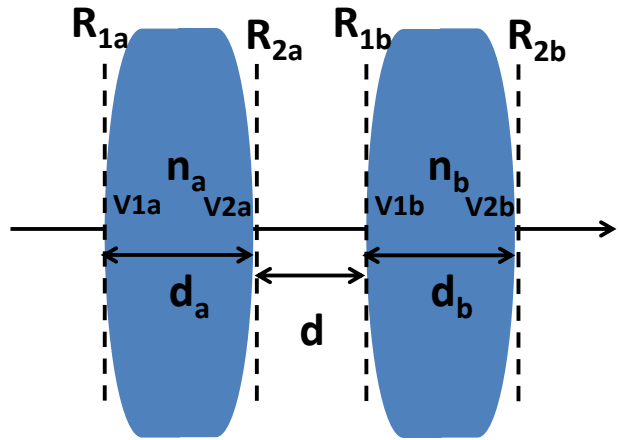

**Figure 11:** Two thick lenses separated a distance '*d*'.

We write, for the matrix of the two lenses system:

$$M_{ab} = \begin{pmatrix} b_{11} & b_{12} \\ b_{21} & b_{22} \end{pmatrix} \begin{pmatrix} 1 & 0 \\ d & 1 \end{pmatrix} \begin{pmatrix} a_{11} & a_{12} \\ a_{21} & a_{22} \end{pmatrix}. \tag{55}$$

Where the elements $a_{ij}$ and $b_{ij}$ are specified in the Eqs. (47) to (54).
The resultant matrix has elements:

$$M_{ab11} = b_{11}a_{11} + b_{12}(da_{11} + a_{21}). \tag{56}$$

$$M_{ab12} = b_{11}a_{12} + b_{12}(da_{12} + a_{22}). \tag{57}$$

$$M_{ab21} = b_{21}a_{11} + b_{22}(da_{11} + a_{21}). \tag{58}$$

$$M_{ab22} = b_{21}a_{12} + b_{22}(da_{12} + a_{22}). \tag{59}$$

It may be illustrative to develop the element $M_{ab12}$ of the compound system using the matrix elements explicitly (Eqs. (47) to (54)). As already seen for one single lens, the element "*1-2*" is related to the focal distance. After a little algebra, Eq. (57) becomes:

$$M_{ab12} = -\frac{1}{f_a} - \frac{1}{f_b} + \frac{d}{f_a f_b} + \frac{(n_a - 1)d_a}{n_a R_{1a} f_b} - \frac{(n_b - 1)d_b}{n_b R_{2b} f_a}. \tag{60}$$

It will be shown that this expression corresponds to the negative of the inverse of the focal distance for the two thick lenses, separated by a distance *d*. I call this focal distance, $f_{ab}$. Meanwhile, it is easy to see that, making $d_a = d_b = 0$, i.e., two thin lenses, Eq. (60) reduces to the well-known expression of the (negative of the inverse) focal distance for two thin lenses separated by a distance *d* (where $f_a$ and $f_b$, of course, are reduced to the expressions of focal distances for thin lenses).

We can determinate the back focal length for this system, proceeding in an analogous way as we did for the single thick lens (see Eq. (8)), i.e.:

$$\begin{pmatrix} \alpha_2 \\ y \end{pmatrix} = \begin{pmatrix} 1 & 0 \\ z & 1 \end{pmatrix} \begin{pmatrix} M_{ab11} & M_{ab12} \\ M_{ab21} & M_{ab22} \end{pmatrix} \begin{pmatrix} \alpha_1 \\ y_1 \end{pmatrix}. \tag{61}$$

Performing the product of matrixes in Eq. (51) and writing the expression for *y*:

$$y = (zM_{ab11} + M_{ab21})\alpha_1 + (zM_{ab12} + M_{ab22})y_1. \tag{62}$$

As we already know, for back focal point, $\alpha_1=0$, and we must compute the value of *z* that makes *y* null for all values of $y_1$:

$$z_b = -\frac{M_{ab22}}{M_{ab12}}. \tag{63}$$

Writing $M_{ab12}$ as $-1/f_{ab}$, so:

$$z_b = f_{ab} M_{ab22}. \tag{64}$$

Now, we take a look to the expression for $z_b$ of the single thick lens, Eq. (11). It may be written, in term of the matrix element of the lens, as $z_b = fa_{22}$.

Also, we can find the front focal length for the compound system of thick lenses, using a reasoning analogous to that used in Eqs. (12) to (15), the result is:

$$z_f = f_{ab} M_{ab11}. \tag{65}$$

Once more, we observe that the expression is formally identic to the corresponding front focal length for a single thick lens, Eq. (15), that may be written using the matrix element as $z_f = fa_{11}$.

The next logical step is to find the principal planes for the two thick lens system and find the relation with the focal distance, $f_{ab}$. See Fig. 12.

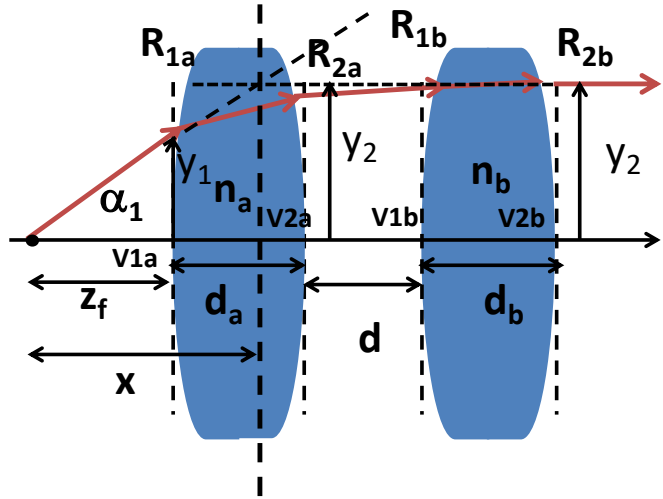

**Figure 12:** Finding principal planes for two thick lenses separated a distance '$d$'.

In the Fig. 12 above, we find the 1PP for the two thick lens system using, as before, the property of front focal point, $z_f$, and projecting back the $y_2$ coordinate. Then write for the angle $\alpha_1$ (in the paraxial approximation):

$$\alpha_1 \cong \frac{y_2}{x} \cong \frac{y_1}{z_f}. \tag{66}$$

Then, we can write for x:

$$x = \frac{y_2}{y_1} z_f. \tag{67}$$

Which is, of course, formally the same equation obtained when the single thick lens was studied. Now, in order to obtain an expression for $y_2$ in our two thick lenses systems, we write:

$$\begin{pmatrix} \alpha_2 \\ y_2 \end{pmatrix} = \begin{pmatrix} M_{ab11} & M_{ab12} \\ M_{ab21} & M_{ab22} \end{pmatrix} \begin{pmatrix} \alpha_1 \\ y_1 \end{pmatrix}. \tag{68}$$

Note that this expression allows us to obtain the height and angle, α₂ and y₂, of a ray leaving the compound system at the last surface (R₂b), as a function of the corresponding parameters α₁ and y₁ of a ray hitting the systems at the first surface (R₁a).

From Eq. (68) we obtain, for y₂:

$$y_2 = M_{ab21}\alpha_1 + M_{ab22}y_1. \tag{69}$$

Substituting α₁ from Eq. (66) in Eq. (69):

$$y_2 = M_{ab21}\frac{y_1}{z_f} + M_{ab22}y_1. \tag{70}$$

From which is trivial obtain:

$$\frac{y_2}{y_1} = \frac{M_{ab21}}{z_f} + M_{ab22}. \tag{71}$$

Now we can substitute Eq. (71) in Eq. (67) and obtain (remember that $z_f = f_{ab}M_{ab11}$):

$$x = M_{ab21} + f_{ab}M_{ab11}M_{ab22}. \tag{72}$$

Our next task is to demonstrate that $x = f_{ab}$. It is a long algebra but also, one of the principal points of this work, so it will be developed here. We write Eq. (72) explicitly as a function of the matrix elements:

$$x = b_{21}a_{11} + b_{22}(da_{11} + a_{21}) + f_{ab}[b_{11}a_{11} + b_{12}(da_{11} + a_{21})][b_{21}a_{12} + b_{22}(da_{12} + a_{22})]. \tag{73}$$

Doing some basic algebra, Eq. (73) can be written as:

$$x = b_{21}a_{11} + b_{22}da_{11} + b_{22}a_{21} + f_{ab}[b_{11}a_{11}b_{21}a_{12} + a_{11}db_{22}(b_{11}a_{12} + b_{12}da_{12} + b_{12}a_{22}) + b_{11}a_{11}b_{22}a_{22} + b_{12}da_{11}b_{21}a_{12} + +b_{12}a_{21}b_{21}a_{12} + b_{12}a_{21}b_{22}da_{12} + b_{12}a_{21}b_{22}a_{22}] \tag{74}$$

We see, in Eq. (74), that the term within the parentheses, multiplying the $a_{11}db_{22}$ factor, is the $M_{ab12}$ matrix element (Eq. (57)), which is equal to -1/ $f_{ab}$. We write then:

$$x = b_{21}a_{11} + b_{22}da_{11} + b_{22}a_{21} + f_{ab}(b_{11}a_{11}b_{21}a_{12} - a_{11}\frac{d}{f_{ab}}b_{22} + b_{11}a_{11}b_{22}a_{22} + b_{12}da_{11}b_{21}a_{12} + b_{12}a_{21}b_{21}a_{12} + b_{12}a_{21}b_{22}da_{12} + b_{12}a_{21}b_{22}a_{22}). \tag{75}$$

In Eq. (75), it is possible now to cancel the second and the fifth terms, doing this and after a rearrangement, we obtain:

$$x = b_{21}a_{11}[1 + f_{ab}(b_{11}a_{12} + b_{12}da_{12})] + b_{22}a_{21} + f_{ab}[b_{11}a_{11}b_{22}a_{22} + b_{12}a_{21}b_{21}a_{12} + a_{21}b_{22}b_{12}(da_{12} + a_{22})]. \tag{76}$$

We can express the factors between parenthesis in the second and sixth terms using the fact that:
-1/$f_{ab}$=$b_{11}a_{12}$ +$b_{12}(da_{12}+ a_{22})$, then:

$$x = b_{21}a_{11}[1+f_{ab}(-\frac{1}{f_{ab}}-b_{12}a_{22})]+b_{22}a_{21}+f_{ab}[b_{11}a_{11}b_{22}a_{22}+b_{12}a_{21}b_{21}a_{12}+a_{21}b_{22}(-\frac{1}{f_{ab}}-b_{11}a_{12})] \qquad (77)$$

After that, it can be obtained:

$$x = f_{ab}b_{12}b_{21}a_{21}a_{12}-f_{ab}b_{12}b_{21}a_{11}a_{22}+f_{ab}b_{11}a_{11}b_{22}a_{22}-f_{ab}b_{11}a_{12}a_{21}b_{22} \qquad (78)$$

And, finally:

$$x = f_{ab}(a_{21}a_{12}-a_{11}a_{22})(b_{12}b_{21}-b_{11}b_{22}). \qquad (79)$$

Now, it must be replaced the elements $a_{ij}$ and $b_{ij}$ explicitly in Eq. (79), using Eqs. (47) to (54). Working with the first factor:

$$a_{21}a_{12}-a_{11}a_{22} = -\frac{d_a}{n_a}\frac{1}{f_a}-\left(1+\frac{n_a-1}{n_a}\frac{d_a}{R_{2a}}\right)\left(1-\frac{n_a-1}{n_a}\frac{d_a}{R_{1a}}\right). \qquad (80)$$

After a simple algebra, it is obtained:

$$a_{21}a_{12}-a_{11}a_{22} = -1-\frac{d_a}{n_a}\frac{1}{f_a}+\frac{d_a}{n_a}\left(\frac{n_a-1}{R_{1a}}-\frac{n_a-1}{R_{2a}}+\frac{(n_a-1)^2}{n_a}\frac{d_a}{R_{2a}R_{1a}}\right). \qquad (81)$$

The term between parentheses, in Eq. (81) is equal to $1/f_a$ (see Eq. (48)), so:

$$a_{21}a_{12}-a_{11}a_{22} = -1. \qquad (82)$$

Obviously, $b_{21}b_{12}$ - $b_{11}b_{22}$ = -1 also. With this, Eq. (79) can be finally written as:

$$x = f_{ab}. \qquad (83)$$

With this demonstration, it has been shown that the interpretation of $f_{ab}$ as the focal distance of the two thick lenses systems is correct, and totally analogous to the case of one single thick lens. We can now write the expression for focal distance of two thick lenses separated a distance $d$ as:

$$-\frac{1}{f_{ab}} = -\frac{1}{f_a}-\frac{1}{f_b}+\frac{d}{f_af_b}+\frac{(n_a-1)d_a}{n_aR_{1a}f_b}-\frac{(n_b-1)d_b}{n_bR_{2b}f_a}. \qquad (84)$$

The principal plane found and shown in Fig. 12 is the first principal plane. Of course, the second principal plane is easily found and is verified that its distance to the back focal point, $z_b$, is also $f_{ab}$

The next steps now are, in close analogy with the single thick lens case, to find the distances of the 1PP and 2PP, $h_1$ and $h_2$, which will be measured with respect to the outermost vertices points of the system, i.e., $V_{1a}$ e $V_{2b}$ respectively, and also determinate the image condition.

In order to find the distances of the principal planes respect to the outermost vertices points, we write:

$$h_1 = f_{ab} - z_f = f_{ab}(1 - M_{ab11}). \tag{85}$$

$$h_2 = z_b - f_{ab} = f_{ab}(M_{ab22} - 1). \tag{86}$$

Where Eqs. (64) and (65) were used to express front and back focal distances for the two thick lens system. As before, $h_1$ and $h_2$ follow the signal convention for object and images, respectively.

To find image condition, we measure the object distance, $s_o$, with respect to $V_{1a}$ and image distance, $s_i$, with respect to $V_{2b}$, see Fig. 13:

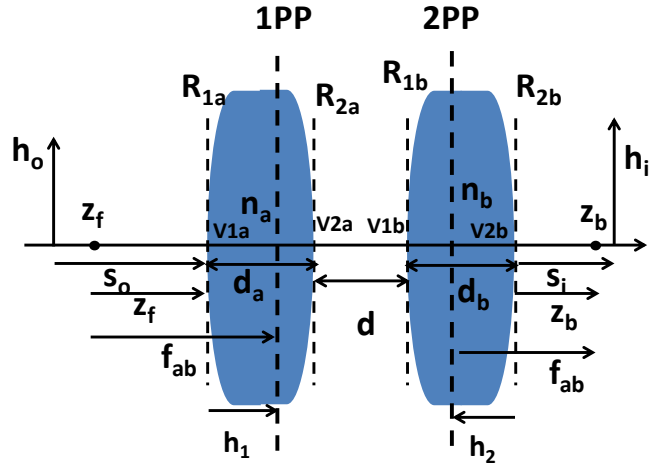

**Figure 13:** Image condition for two thick lenses. Principal planes are located respect to outermost vertices points, $V_{1a}$ and $V_{2b}$.

Writing the matrix equation for the object-two thick lenses system-image:

$$\begin{pmatrix} \alpha_i \\ h_i \end{pmatrix} = \begin{pmatrix} 1 & 0 \\ s_i & 1 \end{pmatrix} \begin{pmatrix} M_{ab11} & M_{ab12} \\ M_{ab21} & M_{ab22} \end{pmatrix} \begin{pmatrix} 1 & 0 \\ s_o & 1 \end{pmatrix} \begin{pmatrix} \alpha_o \\ h_o \end{pmatrix}. \tag{87}$$

We see that Eq. (87) is formally identic to Eq. (26), thus, it is verified that the image condition is:

$$s_i M_{ab11} + M_{ab21} s_i s_o + M_{ab21} + s_o M_{ab22} = 0. \tag{88}$$

Which is formally identic to Eq. (35), therefore, we can now follow the same steps of section 4, Eqs. (36) to (45), to the extent of writing the image condition for the two thick lenses system in the simple form of a Gauss equation. To achieve this, of course, object and image distances must be measured with respect to the principal planes. With this condition, a complicated system as the shown in Fig. 13 is reduced to the one shown in Fig. 14:

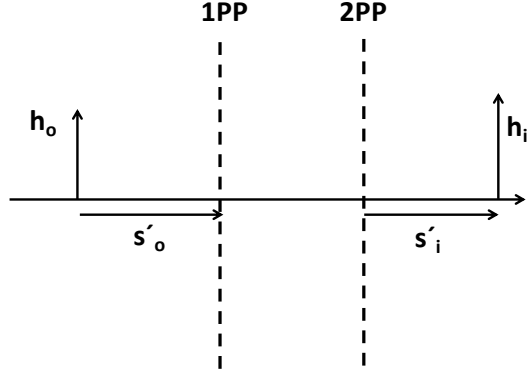

**Figure 14:** Object and Image coordinates for two thick lenses with respect to the principal planes.

Furthermore, that system is described analytically by:

$$\frac{1}{s'_o} + \frac{1}{s'_i} = \frac{1}{f_{ab}}. \tag{89}$$

## VI. THREE AND N THICK LENSES, IMAGE CONDITION.

The expressions for three, four and *N* thick lenses systems follow the same general structure. Consider a three thick lenses system, *A*, *B*, and *C*, being the a, b lenses identical to the system seen in the last section, and *C* lens is located to a distance $d_{bc}$ to the right of *B* lens. The *c* lens, of course has matrix elements mathematically identic to those represented in the Eqs. (47) to (50), and characterized with usual parameters, refraction index $n_c$, thickness $d_c$, radius $R_{1c}$, and $R_{2c}$. The matrix for this system is:

$$M_{abc} = \begin{pmatrix} c_{11} & c_{12} \\ c_{21} & c_{22} \end{pmatrix} \begin{pmatrix} 1 & 0 \\ d_{bc} & 1 \end{pmatrix} \begin{pmatrix} M_{ab11} & M_{ab12} \\ M_{ab21} & M_{ab22} \end{pmatrix}. \tag{90}$$

And the matrix elements:

$$\begin{aligned}
M_{abc11} &= c_{11} M_{ab11} + c_{12}(d_{bc} M_{ab11} + M_{ab21}) \\
M_{abc12} &= c_{11} M_{ab12} + c_{12}(d_{bc} M_{ab12} + M_{ab22}) \\
M_{abc21} &= c_{21} M_{ab11} + c_{22}(d_{bc} M_{ab11} + M_{ab21}) \\
M_{abc22} &= c_{21} M_{ab12} + c_{22}(d_{bc} M_{ab12} + M_{ab22})
\end{aligned} \tag{91}$$

All the characteristics parameters for this system, as the front focal point, back focal point, etcs, can be obtained from these elements, using the expressions already seen for the two thick lenses system, and substituting the corresponding matrix element by the corresponding to the three lenses system. Of course, the focal distance is associated to the $M_{abc12}$ element as usual. A closer look to Eqs (91) allow us to write them in a compact form:

$$M_{abc,ij} = c_{i1} M_{ab,1j} + c_{i2}(d_{bc} M_{ab,1j} + M_{ab,2j}). \tag{92}$$

Where *ij* represents the index of the matrix.
Finally, we can write the expression for the matrix elements of a system of *N* thick lenses:

$$M_{ab\ldots n,ij} = n_{i1} M_{ab\ldots n-1,1j} + n_{i2}(d_{n-1n} M_{ab\ldots n-1,1j} + M_{ab\ldots n-1,2j}). \tag{93}$$

Where $n_{ij}$ represents the matrix elements corresponding to the last thick lens of the system (i.e., the last on the right), and $d_{n-1n}$, the distance between this last lens and the one immediately at left, *N-1*. Of course, the other distances are contained in the $M_{ab...n-1,ij}$ matrix.

As before, we can use the matrix elements of Eq. (93) and the general expressions already derived, to find any parameter of interest of an optical system formed by *N* thick lenses. Also, we can determinate the two principal planes for this system and find image distances with a simple Gaussian equation (Eq. (89)), keeping in mind that the focal distance is always related to the $M_{ab..n,12}$ element and the object an image distances should be measures with respect to the 1PP and 2PP, respectively.

**VII. CONCLUSIONS.**

In this work, I have studied thick lenses immersed in air using the matrix formalism in the paraxial approximation. It has been determined their characteristics parameters, such as focal distance, back and front focal points, principal planes and object-image equation. A simple Gaussian equation, identic to the used for the idealized case of thin lenses is found to relate the object and image distances, when these are measured with respect to the principal planes. Also, the analysis has been extended to two thick lenses separated a distance *d*. It has been determined an expression for the focal distance of this system, and also the back and front focal points and principal planes. It was settled that the mathematical expressions determining these parameters are formally identic to the corresponding to the single thick lens, as a function of the matrix elements characterizing both systems. Also for the case of two thick lenses, once determined the principal planes, it was found that a simple Gaussian equation relates the object and image distances measured with respect to the principal planes. Finally, extending this analysis to three and *N* thick lenses systems, analogous results was found.

It is interesting to observe that all well-known expressions relating thin lenses systems can be deduced from the more realistic formulas derived in this work, simply by making the parameter $d_a$ of the lens equal to zero, i.e., passing from a thick to a thin lens.

It is of particular interest the result that, given an optical system with an arbitrary number of lenses, allow us to find a single pair of principal planes associated to that system, and determinate image distances (given the object distances) by resolving a simple Gaussian equation. This last result, that was mentioned by Feynman in his lectures book, although well known to all specialist in the field, had not been formally derived in the basic or intermediate optic literature to the best of my knowledge.

**References**

1. Hecht E. *Optics* 5nd edition (Pearson, 2017), p.256.
2. Jenkins F., White H., Fundamental of Optics, 4nd edition (McGraw-Hill Education, 2001), p.66.
3. Born M., Wolf E., Principles of Optics, 7nd edition (Cambridge University Press, 1999), p.173.
4. Feynman, R. P., The Feynman lectures on physics, definitive edition (Pearson-Addison-Wesley, California, 2006). Volume I, p. 27-6.

**APPENDIX**

Our task here is to demonstrate that the following identity is true:

$$a_{12}oi + a_{21} + a_{22}o + ia_{11} = 0. \tag{A1}$$

Where *o* and *i* are given by Eqs. (42) and (43). Substituting then those equations in Eq. (A1) we have:

$$a_{12}\left(\frac{1-a_{11}}{a_{12}}\right)\left(\frac{1-a_{22}}{a_{12}}\right) + a_{21} + a_{22}\left(\frac{1-a_{11}}{a_{12}}\right) + \left(\frac{1-a_{22}}{a_{12}}\right)a_{11} = 0. \tag{A2}$$

After the basic algebra and further simplifications, Eq. (A2) is reduced to:

$$\frac{1}{a_{12}} + a_{21} - \frac{a_{22}a_{11}}{a_{12}} = 0. \tag{A3}$$

Substituting the expressions for the matrix elements $a_{ij}$, Eq. (A3) becomes:

$$-f_a + \frac{d_a}{n_a} + f_a\left(1 - (n_a - 1)\frac{d_a}{n_a R_{1a}}\right)\left(1 + \frac{n_a - 1}{n_a}\frac{d_a}{R_{2a}}\right) = 0. \tag{A4}$$

After the usual algebra:

$$f_a \frac{d_a}{n_a}\left[(n_a - 1)\left(\frac{1}{R_{2a}} - \frac{1}{R_{1a}}\right) - \frac{(n_a - 1)^2}{n_a}\frac{d_a}{R_{1a}R_{2a}}\right] + \frac{d_a}{n_a} = 0. \tag{A5}$$

Clearly, the term between brackets in Eq. (A5) is -1/$f_a$, and the identity is verified.